\documentclass[12pt]{iopart}

\usepackage{graphicx}
\usepackage[small,bf]{caption}
\begin{document}
\title[Two- and three-particle correlations of high-$p_{t}$ charged hadrons at 158A GeV/c]{Two- and three-particle azimuthal correlations of high-$\mathbf p_{\mathbf t}$ charged hadrons in Pb-Au collisions at \\ 158A GeV/c }
\author{Stefan Kniege and Mateusz Ploskon \footnotesize{for the CERES collaboration}}
\address{ Institut f\"ur Kernphysik, Max-von-Laue-Str. 1, 60438 Frankfurt am Main, Germany}
\ead{kniege@ikf.uni-frankfurt.de, ploskon@ikf.uni-frankfurt.de}
\begin{abstract}
Azimuthal correlations of hadrons with high transverse
momenta serve as a measure to study the energy loss and the 
fragmentation pattern of jets emerging from hard parton-parton interactions in heavy ion collisions.
Preliminary results from the CERES experiment on two- and three-particle correlations
in central Pb-Au collisions are presented. A strongly non-Gaussian shape on the away-side of the
two-particle correlation function is observed,
indicating significant interactions of the emerging partons with the medium.
Mechanisms like deflection of the initial partons
or the evolution of a mach cone in the medium can lead to similar 
modifications of the jet structure on the away-side. An analysis based 
on three-particle correlations is presented which helps to shed light on the origin of 
the observed away-side pattern.
\end{abstract}
\section{Introduction}
The suppression of high-$p_{t}$ charged hadrons in A-A relative to p-p collisions 
observed at RHIC \cite{STAR,PHENIXsup} was assumed to arise from the interaction of partons emerging 
from hard collisions within a dense colored medium \cite{GYULhijing}.   
In such a picture hard collisions close to the surface of the reaction zone could lead to a 
jet fragmenting into the vacuum while the second jet would have to traverse the medium 
suffering energy loss and hence redistributing part of its energy to the medium.
Measuring azimuthal angular correlations between a high-$p_t$ trigger particle and associated 
particles revealed differences in the shape of the correlation function
close to the trigger particle (near-side) and around $\Delta$$\phi$~=~$\pi$ (away-side). Depending on the momentum of the associated and trigger particles a strong suppression of the yield on the 
away-side of the correlation function was observed \cite{STARDissapp,JANA,Adams:2005ph,PHENIX} as compared to p-p data. 
In addition, the away-side exhibited a double peak structure which was proposed to arise from
Mach cone shock waves \cite{STOECKER}, induced gluon radiation \cite{VITEVcher} or Cherenkov-like radiation in the medium \cite{Koch}.
Those scenarios can not be distinguished from an event-by-event deflection of the jets in the medium based on 
two-particle correlations only.
Three-particle correlation measurements, i.e. measurements of correlations among two associated particles in an event 
with respect to a trigger particle can help to disentagle the different proposed scenarios. 
The two-particle analysis follows previous measurements of non-triggered azimuthal correlations performed 
by the CERES collaboration \cite{JANA} which already indicated modifications on the away-side of the di-jet 
correlation function in Pb-Au collisions at SPS energy. 
The present analysis is based on 30 million Pb-Au events at 158A GeV/c recorded with 
the upgraded CERES spectrometer including a Time Projection Chamber covering the full range 
in azimuthal angle, and $2.1<\eta<2.7$. 
\section{Two-particle correlations}
The (jet-) associated particles are investigated by measuring the distribution $S(\Delta\phi)$ of the difference 
in the azimuthal angle $\Delta\phi$ of a trigger particle with $2.5<p_{t}<4.0$ GeV/c and all associated particles
with $1.0< p_{t} < 2.5$ GeV/c from the same event.
To account for acceptance effects, the signal distribution is divided by 
a background distribution $B(\Delta\phi)$ where trigger and associated particles are taken from different events.
The correlation function $C_{2}(\Delta\phi)=\frac{n_{backgr}}{n_{sig}}\cdot \frac{S(\Delta\phi)}{B(\Delta\phi)}$, 
normalized to the ratio of background to signal pairs ($n_{backgr}/n_{sig}$), is assumed to be composed of two components: The jet-like correlations, and the correlations due to the elliptic flow. 
In this two-source approach the measured correlation function can be written in the form:\\
\hspace*{3.0cm} $C_{2}(\Delta\phi)=C_{2,jet}(\Delta\phi)+b\cdot(1+2 < v_2^T v_2^A > \cos(2 \Delta \phi)).$ \\
The elliptic flow coefficients for the trigger ($v_2^T$) and associated ($v_2^A$) range are determined by a reaction plane analysis method \cite{JANA,JOVAN}.
Assuming Zero Yield At Minimum (ZYAM) \cite{PHENIX} for the jet-like correlation, $C_{2,jet}$, the elliptic flow contribution is determined and subtracted from the correlation function. 
After normalisation we obtain the conditional yield as the number of jet-associated particles per trigger $N^T$:
$\hat{J_{2}}(\Delta\phi)=\frac{1}{N_{T}}\frac{dN^{TA}}{d\Delta \phi} = \frac{C_{2,jet}(\Delta \phi)}{\int(C_2(\Delta \phi ')d(\Delta \phi '))} \frac{N^{TA}}{N^{T}}.$
\begin{figure}[h!]
\centering
\begin{minipage}[t]{15.8cm}
\centering
\includegraphics[width=1.0\textwidth]{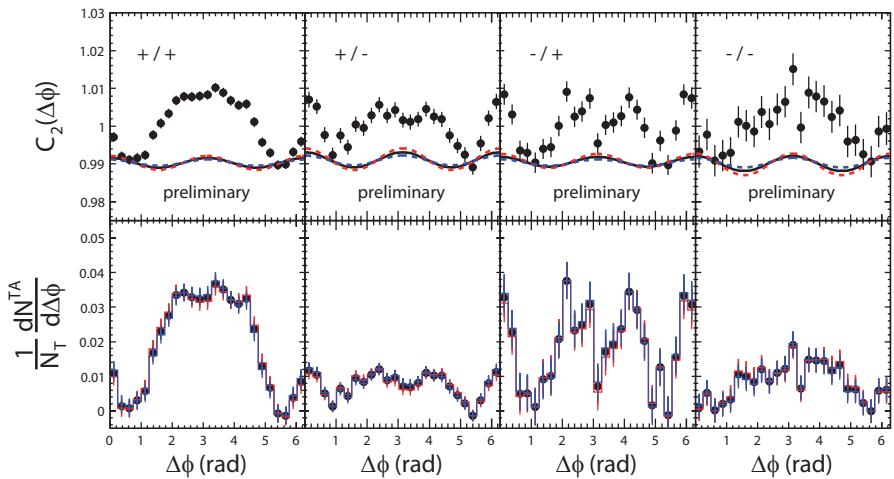}
\caption{ {\footnotesize Two-particle correlation function (upper row) and conditional yield (lower row) for different 
carge combinations ( trigger charge/associated charge ) in the centrality range
$\sigma/\sigma_{geom}$= (0-5)\%. The flow contribution is indicated by the black lines in the upper row. 
The dashed and dotted curves show the systematic uncertainties in the flow contribution  
($\sigma_{<v_2^A>}\approx 15\%\ , \sigma_{<v_2^T>}\approx 25\% $) . The effect of the flow uncertainty 
on the conditional yield is shown by the blue and red line (lower row) (color online).} }
\label{fig:2part}
\end{minipage} 
\end{figure}
The upper row of Figure \ref{fig:2part} shows the correlation functions $C_2(\Delta\phi)$ 
and the estimated flow contributions for different charge combinations of trigger and 
associated particles. The lower row shows the extracted conditional yield corrected for 
efficiency which was estimated to be $85\%$ for momenta above 1 GeV/c. 
As reported earlier \cite{JANA,MATEUSZqm} a non-Gaussian shape is observed on the
away-side, indicating significant medium modifications in central Pb-Au 
collisions at SPS energy. The observed shape and relative magnitude of 
near- and away-side depend on the trigger and associated particle charges. 
For a given trigger charge, the associated yield on the near side is larger
for unlike-sign combinations as compared to like-sign. This is qualitatively
explained by local charge conservation in the fragmentation process. 
The trend is different on the away-side: For both trigger charges,
the yield is significantly larger for positive associated particles 
compared to negative associates.
PYTHIA \cite{PYTHIA} calculations at $\sqrt{s_{\rm NN}}=17.3$~GeV
indicate isospin-dependent effects when comparing positive and negative
associated yields in pp, pn and nn collisions. This situation arises due to the 
dominant contribution of large-$x$ (valence quark) scattering to the di-jet yield around
mid-rapidity at SPS energy.
However, a full account of the observed charge asymmetry could not be
achieved based on PYTHIA calculations.
It should be noted that PYTHIA shows little or no charge asymmetries at RHIC and LHC
due to the increasing contribution from gluons.
For all charge combinations, a significant broadening of the away-side peak compared
to the expectation in pp is observed, perhaps even indicating the occurance of
a slight dip at $\Delta \phi = \pi$ in the case of unlike-sign combinations. 
\section{Three-particle correlations}
\begin{figure}[h!]
\centering
\begin{minipage}[t]{15.8cm}
\centering
\includegraphics[width=1.0\textwidth]{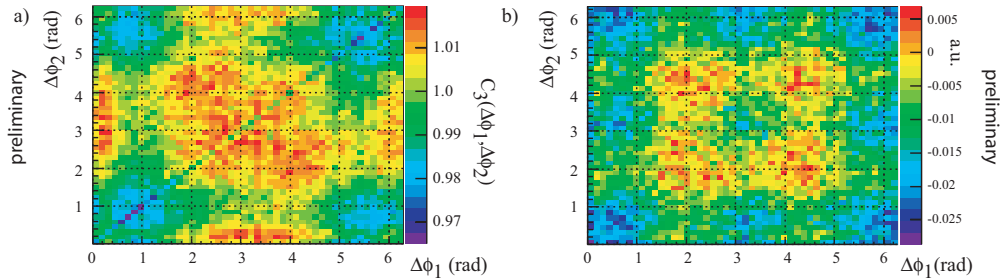}
\caption{ {\footnotesize Three-particle correlation function a) and jet-like three-particle yield after background subtraction b) for all charge combinations in central Pb-Au collisions at 158A GeV/c. }}
\label{fig:3part}
\end{minipage} 
\end{figure}

To analyze the shape on the away-side the differences in the azimuthal angle for 
two associated particles with respect to a trigger particle  $(\Delta\phi_{1},\Delta\phi_{2})$ 
are acquired to obtain the three-particle signal distribution $J_3(\Delta\phi_{1},\Delta\phi_{2})$.
The same $p_t$-intervals for trigger and associated particles are used as for the two-particle analysis. 
Dividing the signal by a normalized background distribution where all three particles are taken from
different events the three particle correlation function $C_3(\Delta\phi_1,\Delta\phi_2)$ is obtained 
as depicted in Figure \ref{fig:3part}(a).
The signal distribution is composed of the genuine three particle jet yield $\hat{J_{3}}(\Delta\phi_{1},\Delta\phi_{2})$ and several background components which are subtracted from the signal as described in detail in \cite{ULERY3part}:
$\hat{J_{3}}(\Delta\phi_{1},\Delta\phi_{2})=J_{3}(\Delta\phi_{1},\Delta\phi_{2}) -a\cdot\hat{J_2}\otimes B_{2} - ba^2(B_3^{mb} +B_3^{mb,flow})$.
The hard-soft component $\hat{J_2}\otimes B_{2}$ corresponds to the case when one associated particle
is jet-like correlated to the trigger while the second one is from the background. It is constructed by folding
the 2-particle conditional yield $\hat{J_{2}}$ with the flow modulated background.      
The third term is denoted as soft-soft background and corresponds to the case when the
trigger is a non-jet particle. $B_3^{mb}$ is constructed by mixing two associate particles from 
one event with a trigger particle from another event to account for all correlations 
among the associated particles which are not correlated to the trigger.
The term $B_3^{mb,flow}$ accounts for all three particles being flow-like correlated.
The mixing is based on events without trigger condition labled {\itshape mb} (minimum bias). 
The paramter $a$ accounts for the different multiplicities in triggered and non-triggered events. 
Subtracting the hard-soft background, one is left with the correlations where all particles are background particles or jet-like
correlated. 
Deploying the ZYAM method for the genuine jet-like correlations the soft-soft background is adjusted 
with the parameter $b$ and subtracted to obtain the  jet-correlated yield of the associated 
particles (Figure \ref{fig:3part}(b)).
Clear off-diagonal components are observed indicating a cone-like emission of the associated particles in high-$p_t$ triggered events.
\section[Conclusion]{Conclusion and Outlook}
We presented two- and three-particle correlations for high-$p_t$ charged hadrons at 158A GeV/c indicating 
that already at SPS energy medium effects on high-$p_t$ particles are observed. 
The magnitude and shape of the two-particle jet yield depends on the charge combination of trigger and associated particles.
In addition, a non-Gaussian shape on the away-side is observed which is reflected in 
off-diagonal structures in the three-article jet yield. 
Further investigations are needed to substantiate these findings.

\end{document}